\documentclass[11pt,a4paper]{article}

\usepackage{jheppub}
\usepackage{color}

\newcommand{\nn}{\nonumber}
\def\dfrac#1#2{\displaystyle\frac{#1}{#2}}

\newcommand{\pslash}{p\kern-1ex /}
\newcommand{\qslash}{q\kern-1ex /}
\newcommand{\lslash}{l\kern-1ex /}
\newcommand{\sslash}{s\kern-1ex /}
\newcommand{\kaslash}{k_a\kern-2ex /}
\newcommand{\kbslash}{k_b\kern-2ex /}
\newcommand{\Dslash}{{\cal D}\kern-1.5ex /}

\newcommand{\beqa}{\begin{eqnarray}}
\newcommand{\eeqa}{\end{eqnarray}}

\preprint{YITP-22-102}
\begin{document}
\title{Colliding gravitational waves and singularities}
\author[a]{Sinya AOKI,}
\affiliation[a]{Center for Gravitational Physics and Quantum Information, Yukawa Institute for Theoretical Physics, Kyoto University, Kitashirakawa Oiwakecho, Sakyo-ku, Kyoto 606-8502, Japan}

\abstract{We have investigated a model of colliding plain gravitational waves, proposed by Szekeres,
whose structure of singularities is determined. 
We have evaluated a total energy of matter 
as a volume integral of the energy momentum tensor (EMT), whose contributions arise only at these singularities.
The total matter energy
is conserved before a collision of two plane gravitational waves but decreases during the collision and becomes zero at the end of the collision.
We thus interpret that this model of colliding plane gravitational waves is a spacetime describing a pair annihilation of plan gravitational waves.
We have also calculated a matter conserved charge proposed by the present author and his collaborators.
The matter charge is indeed conserved but is zero due to a cancellation between two plain gravitational waves.
This seems natural since nothing remains after a pair annihilation, and give a hint 
on a physical interpretation of the conserved charge, which we call the gravitational charge.
By modifying the space time for the pair annihilation, we newly construct
two types of a scattering plane gravitational wave and a pair creation of  plane gravitational waves, and combining all, a Minkowski vacuum bottle,
a Minkowski spacetime surrounded by two moving plane gravitational waves.
}

\maketitle

\section{Introduction}
Is a total energy always conserved in general relativity ? To answer this question theoretically is rather non-trivial.  
Usually in a flat spacetime, 
as a consequence of Noether's (first) theorem for a global symmetry\cite{Noether:1918zz},
a time translation symmetry of a system defines a corresponding energy density as the time  component of the Noether current, and also tells us that a total energy given by a volume integral of the energy density is conserved. 
This construction of a conserved energy  does not work in general relativity, however, since a time translational symmetry is a part of local symmetries, general coordinate transformations,
to which Noether's first theorem can not be applied.
Indeed, Noether's second theorem\cite{Noether:1918zz} tells us that, instead of dynamically conserved currents, 
there exist non-trivial constraints in a theory with local symmetries
such as Bianchi identity or Gauss law constraint.
Related but not equivalent to this problem, a conservation of the energy momentum tensor (EMT) in general relativity, $\nabla_\mu (\sqrt{-g} T^{\mu\nu})=0$ does not directly give a conserved energy or momentum due to non-linear terms in covariant derivatives.  
There have been several proposals for a definition of energy in general relativity, which may be categorized into two types.
One is Einstein's pseudo-tensor definition, and the other is a quasi-local definition including Komar energy\cite{Komar:1958wp} or ADM energy\cite{Arnowitt:1962hi}.
Since both definitions correspond to some  constraints implied by Noether's second theorem\cite{Aoki:2022gez},
it is easy to show that their current densities are always conserved without using equations of motion, and moreover, the corresponding total energy can be always written as a surface integral rather than a volume integral via Stoke's theorem.

Recently, the present author, together with his collaborators, has proposed an alternative definition for energy in general relativity\cite{Aoki:2020prb,Aoki:2020nzm} without using constraints from Noether's second theorem.
In a case that a stationary Killing vector exists,
the energy defined from the EMT\footnote{In this case, this energy agrees with the one proposed in Ref.~\cite{Fock1959}.}
correctly reproduces the known mass of the Schwarzschild black hole,
and the mass and the angular momentum of the BTZ black hole\cite{Aoki:2020prb}.
For a spherically symmetric compact star, on the other hand, the energy in our definition is different from the corresponding ADM mass,
and its difference may be interpreted as a positive contribution of gravitational fields to the ADM mass. 
Even in the absence of   the stationary Killing vector, the energy is shown to be conserved  during some types of gravitational collapses\cite{Aoki:2020nzm,Yokoyama:2021nnw}.
Since the energy is defined from the EMT of matters, we may call it a ``matter energy" hereafter. 
In general, however, the matter energy is not conserved, so that an answer to the question in the beginning is ``No.".
Although the matter energy is not conserved, one can still define a more general conserved charge associated with the EMT in general relativity,\footnote{This definition is a generalization of the conserved charge proposed for a spherically symmetric spacetime in Ref.~\cite{Kodama:1979vn}. }  
which behaves like ``entropy" for special cases such as homogeneous and isotropic expanding universe\cite{Aoki:2020nzm}.  

In this paper, we apply the definition of matter energy and its generalization in general relativity to a non-trivial dynamical processes that
two plane gravitational  waves collides with each other and  become a black hole-like object, 
which is a vacuum solution to the Einstein equation except singularities. 
(See Refs.~\cite{Szekeres:1972uu,Khan:1971vh,Yurtsever:1988ks,Yurtsever:1988vc} for early analytic studies on this type of processes.)
We exclusively consider a model of colliding plane gravitational waves proposed by Szekeres\cite{Szekeres:1972uu},
which gives a simple analytic solution to the vacuum Einstein equation with some singularities. 
Although it turns out that singularities after the collision is not a black hole but something similar,
this model is still suitable to check whether our proposal  for  matter energy and its generalization works well even for this complicated dynamical processes. 
In Sec.~\ref{sec:PGW}, we explain the model in detail and determine a spacetime structure including singularities.
We then calculate the total matter energy of the system  in Sec.~\ref{sec:charge}, as a volume integral of the EMT, whose non-zero contributions
comes form its singularities.
It turns out that the total matter energy is not conserved during the collision,
though it is conserved before the collision.
On the other hand, the generalized matter charge, which we call the gravitational charge,  can be defined so as to be conserved but is found to be zero by a cancellation between two contributions.
Indeed a geometry of the Szekeres' model is found to describe a pair annihilation of two plane gravitation waves at a certain time, when a total matter energy also becomes zero. The vanishing gravitational charge is consistent with this geometry. 
 In Sec.~\ref{sec:New}, modifying a geometry of the colliding two gravitational waves, we construct several new spacetime geometries, which are
two types of a scattering plane gravitational wave, and a creation of two plane gravitational wave from nothing.
Combining all four, a pair creation of two plane gravitational wave, a scattering of each plane gravitational wave, and a pair annihilation of them,
we finally construct a geometry, named ``Minkowski vacuum bottle", a Minkowski spacetime surrounded by two plane gravitational waves.
Our conclusions, together with some discussions, are given in Sec.~\ref{sec:Concl}.
In appendix~\ref{app:Komar}, we evaluate the generalized Komar integral as a representative of quasi-local definitions of energy for the the Szekeres' model
for a comparison.

\section{Colliding plane gravitational waves}
\label{sec:PGW}

\subsection{Setup}
We consider colliding plane gravitational waves, investigated by Szekeres\cite{Szekeres:1972uu},
whose metric is given by
\beqa
ds^2 &=& - 2e^{-M(u,v)} du dv + e^{-U(u,v)} \left( e^{V(u,v)} (dy^1)^2 + e^{-V(u,v)} (dy^2)^2 \right) , 
\label{eq:metric_rosen}
\eeqa  
where $u=(\tau+z)/\sqrt{2}$ and $v=(\tau-z)/\sqrt{2}$ are light-cone coordinate.
In a flat spacetime, $\tau$ corresponds to a time, while $(x,y,z)$ denotes a spacial position in the Cartesian coordinate.
The {\it vacuum} Einstein equation reads
\beqa
G_{uu}&:=& U_{uu} +M_u U_u -{1\over 2}(U_u^2+V_u^2) = 0, \\
G_{vv} &:=& U_{vv} +M_v U_v -{1\over 2}(U_v^2+V_v^2) = 0, \\
G_{uv}&:=& -U_{uv} + U_u U_v = 0, \\
G^{y^1}{}_{y^1} &:=& e^M\left[ (M+U+V)_{uv} -{1\over 2}(U+V)_u (U+V)_v\right]=0, \\
G^{y^2}{}_{y^2} &:=& e^M\left[ (M+U-V)_{uv} -{1\over 2}(U-V)_u (U-V)_v\right]=0, 
\eeqa
where subscript $u,v$ represent derivatives with respect to them.

\subsection{Solutions to the {\it vacuum} Einstein equation}
Ref.~\cite{Szekeres:1972uu} has solved the above Einstein equations in the 4 regions defined by
\begin{enumerate}
\item [I.] Minkowski spacetime at $u<0$ and $v<0$.
\item[II.]  A left moving plane wave at $u>0$ and $v<0$.
\item[III.]  A right moving plane wave at $u<0$ and $v>0$.
\item[IV.]  Colliding two plane waves  at $u>0$ and $v>0$.
\end{enumerate}

A class of the explicit solutions is given by\cite{Szekeres:1972uu} 
\beqa
U &=& -\log t^2, \ V=-{k_{12}\over 2} \log t^2 + k_1\log (w+p) + k_2\log(r+q), \nn \\
M &=& {1\over 2} \left(1-{k_{12}^2\over 4}\right) \log t^2 +{k_1^2\over 4} \log w + {k_2^2\over 4} \log r +{k_1k_2\over 2} \log(pq+rw), 
\label{eq:sol_IV}
\eeqa
where 
\beqa
p&=&\sqrt{{1\over 2} -f}, \quad  q=\sqrt{{1\over 2} -g}, \quad  r=\sqrt{{1\over 2} + f}, \quad  w=\sqrt{{1\over 2} + g}, \quad  t=\sqrt{f+g},
\label{eq:pqrw}\\
f(u) &=&{1\over 2} -(a u)^{n_1} \theta(u), \quad g(v)={1\over 2} - (bv)^{n_2}\theta(v),
\label{eq:fg}
\eeqa
$k_{12}:= k_1+k_2$, and
$k_1,k_2$ are related to even integers $n_1,n_2$ as
\beqa
{k_i^2 \over 8} &=& 1-{1\over n_i}, \quad i=1,2,
\eeqa
These relations lead to
\beqa
{f_{uu}\over f_u^2} = -{k_1^2\over 8p^2}, \quad {g_{vv}\over g_v^2} = -{k_2^2\over 8 q^2}.
\eeqa

Thanks to $\theta$ functions in \eqref{eq:fg}, these solutions hold in all regions.
As a continuity of the metric and its first derivative,
the strongest requirements become
\beqa
V_u (u) &=& k_1 (n_1a){(ua)^{n_1/2-1}\over 2 r^2} = 0
\eeqa
at the I-II boundary ($u=0$), and 
\beqa
V_v (v) &=& k_2 (n_2b ){(bv)^{n_2/2-1}\over 2 w^2} = 0
\eeqa
at the I-III boundary ($v=0$),
which leads to $ n_1 >2$ and $n_2>2$ for even integers $n_1,n_2$.
Other conditions that  $f=1/2$, $M=V=W=f_u=M_u=0$ at $u=0$ and  $g=1/2$, $M=V=W=g_v=M_v=0$ at $v=0$ are automatically satisfied.
Note that these continuity conditions exclude two colliding impulse waves\cite{Khan:1971vh} and two colliding sandwich plane waves\cite{Yurtsever:1988ks,Yurtsever:1988vc}, both of which have $n_1=n_2=2$.

\subsection{Singularities}
In the region IV, the Riemann curvature invariant shows a singular behavior at $t^2=f+g=0$ as
\beqa
R_{\mu\nu\alpha\beta} R^{\mu\nu\alpha\beta} &\simeq& 
\frac1{64}\left(k_{12}^2-4\right)^2\left(k_{12}^2+12\right)(n_1a)^2(n_2b)^2 p^{k_1^2}q^{k_2^2}(2pq)^{k_1k_2}\times t^{-6-k_{12}^2/2}~~~~
\label{eq:RR}
\\
&\simeq& a^2b^2 \times\left\{\begin{array}{cc}
       63(2pq)^{8}t^{-14} &(n_1=n_2=2)\\
        900(2pq)^{12}t^{-18} &(n_1=n_2=4)\\
        \cdots
    \end{array}
    \right. 
\eeqa
at $t^2\simeq 0$,  while the invariant identically vanishes in other regions.

In regions II and III, Riemann tensors and Weyl tensors have singularities\cite{Szekeres:1972uu},
which are indeed physical, not coordinate singularities as shown below.
In the region II, the metric in \eqref{eq:metric_rosen} can be transformed to the Brinkmann coordinate\cite{Brinkmann:1923cd} as
\beqa
ds^2 = -2 d\bar u d\bar v + A_{ij} x^i x^j (d\bar u )^2 + \sum_{i=1} (dx^i)^2,
\label{eq:metric_brinkmann}
\eeqa
through the coordinate transformation defined by 
\beqa
\bar u &=& \int^u e^{-M(u^\prime)} du^\prime,\ \bar v = v+\sum_{i=1}^2 {1\over 2e_i} {d e_i\over d\bar u} (x^i)^2, \nn \\
x^1 &=& e_1 y^1, \ x^2= e_2 y^2, \quad
e_1(u) := e^{-(U-V)/2}, \ e_2(u) :=  e^{-(U+V)/2},
\eeqa
 where 
 \beqa
 A_{ij} &=& \delta_{ij} {e^{2M}\over e_i}\left[ {d^2 e_i\over du^2} + M_u {d e_i\over du}\right], 
 \eeqa
 and $R_{\bar u i \bar u j} = - A_{ij}$ are only non-zero components of  the Riemann tensor in this coordinate. 
 Explicitly, we obtain
 \beqa
 A_{11} &=& -A_{22} = {k_1\over 8}\left(1-{2\over n_1}\right) { n_1^2 a^2 (au)^{n_1/2-2}\over (t^2)^{3-2/n_1}},
 \label{eq:AII}
 \eeqa
 which is regular at $u=0$ for $n_1\ge 4$, but singular at $t^2 (= 1 -(au)^{n_1}=r^2) =0$ as $(t^2)^\alpha$ with $ 2< \alpha (= 3-{2\over n_1}) < 3$.
 Similarly 
 \beqa
  A_{11} &=& -A_{22} = {k_2\over 8}\left(1-{2\over n_2}\right) { n_2^2 b^2 (bv)^{n_2/2-2}\over (t^2)^{3-2/n_2}}
  \label{eq:AIII}
\eeqa
in the region III.
According to Ref.~\cite{Blau:2011ab},
singularities in \eqref{eq:AII}  and \eqref{eq:AIII} at $t^2=0$ are physical for plane waves in the regions II and III.
 
 In summary, the metric in \eqref{eq:metric_rosen} has singularities at $t^2=0$ in region II, III and IV.

 \subsection{Apparent horizon}
 Since there are singularities at $t^2=0$, we determine locations of apparent horizons if exist.  
 For this purpose, we calculate an expansion $\Theta$ for an affinely parametrized null geodesic\footnote{We here use $\Theta$ instead of $\theta$ to represent the expansion, in order to avoid a confusion with the step function $\theta(x)$.} , defined by
 $
 \Theta := g^{\mu\nu} \nabla_\nu k_{\mu}
 $,
 where $k^\mu$ is the tangent vector of the affinely parametrized null geodesic satisfying $ k^\nu\nabla_\nu k^\mu=0$ and $k_\mu k^\mu =0$.
 The apparent horizon (AH) is determined by the condition that $\Theta$ changes its sign.
 Since it is easy to see that $\Theta=0$ in the region I (flat spacetime), there is no AH in the region I as it should be.
 
 In the region II,  two tangent vectors for future directed null geodesics are given by
 \beqa
 k_+^\mu=( e^{M(u)},0,0,0), \quad k_-^\mu=( 0,e^{M(u)},0,0).
 \eeqa
 Corresponding expansions $\Theta_\pm$ are calculated as
 \beqa
 \Theta_+ = - e^M U_u =- n_1a (au)^{n_1-1} (t^2)^{1/n_1-3/2}, \quad \Theta_-=0.
 \eeqa 
 Since $ \Theta_+ < 0 $ at $u>0$ and  $ \Theta_+ =0 $ at $u=0$,
 we identify  a boundary between I and II at $u=0$  as  the apparent horizon.
 Note that, even though  $ \Theta_+ $ does not becomes positive but stay zero  at $u< 0$, 
 we here extend a meaning of apparent horizon as a boundary between $\Theta <0$ and $\Theta=0$ regions.

 Similarly in the region III, we have
 \beqa
 \Theta_+=0,  \quad  \Theta_- = - e^M U_v = - n_2b (bv)^{n_2-1} (t^2)^{1/n_2-3/2},
 \eeqa
 for $ k^\mu_+=(e^{M(v)},0, 0,0)$ and $k^\mu_-=( 0,e^{M(v)},0,0)$, so that the apparent horizon appears at $v=0$ ( a boundary between I and III).
 
 In the region IV, we take $ k^\mu_+=(e^{M(u,v)},0, 0,0)$ and $k^\mu_-=( 0,e^{M(u,v)},0,0)$, which lead to
 \beqa
 \Theta_+ &=& -e^M U_u =  {e^M\over t^2} (-n_1 a) (a u)^{n_1-1}, \quad
 \Theta_- =  -e^M U_v =  {e^M\over t^2} (-n_2 b ) (b v)^{n_2-1},
 \eeqa
 where
 \beqa
 e^M &=& {w^{k_1^2\over 4} r^{k_2^2\over 4} (pq+rw)^{k_1k_2\over 2} \over (t^2)^{ {(k_1+k_2)^2\over 8} -{1\over 2}}} .
 \eeqa
 Thus the AH appears at $u=0$ (a boundary between II and IV) for $\Theta_+$ and at $v=0$ (a boundary between III and IV) for $\Theta_-$.
 
\subsection{Global spacetime structure}
 \begin{figure}[htb]
\begin{center}
\includegraphics[width=0.8\textwidth,clip]{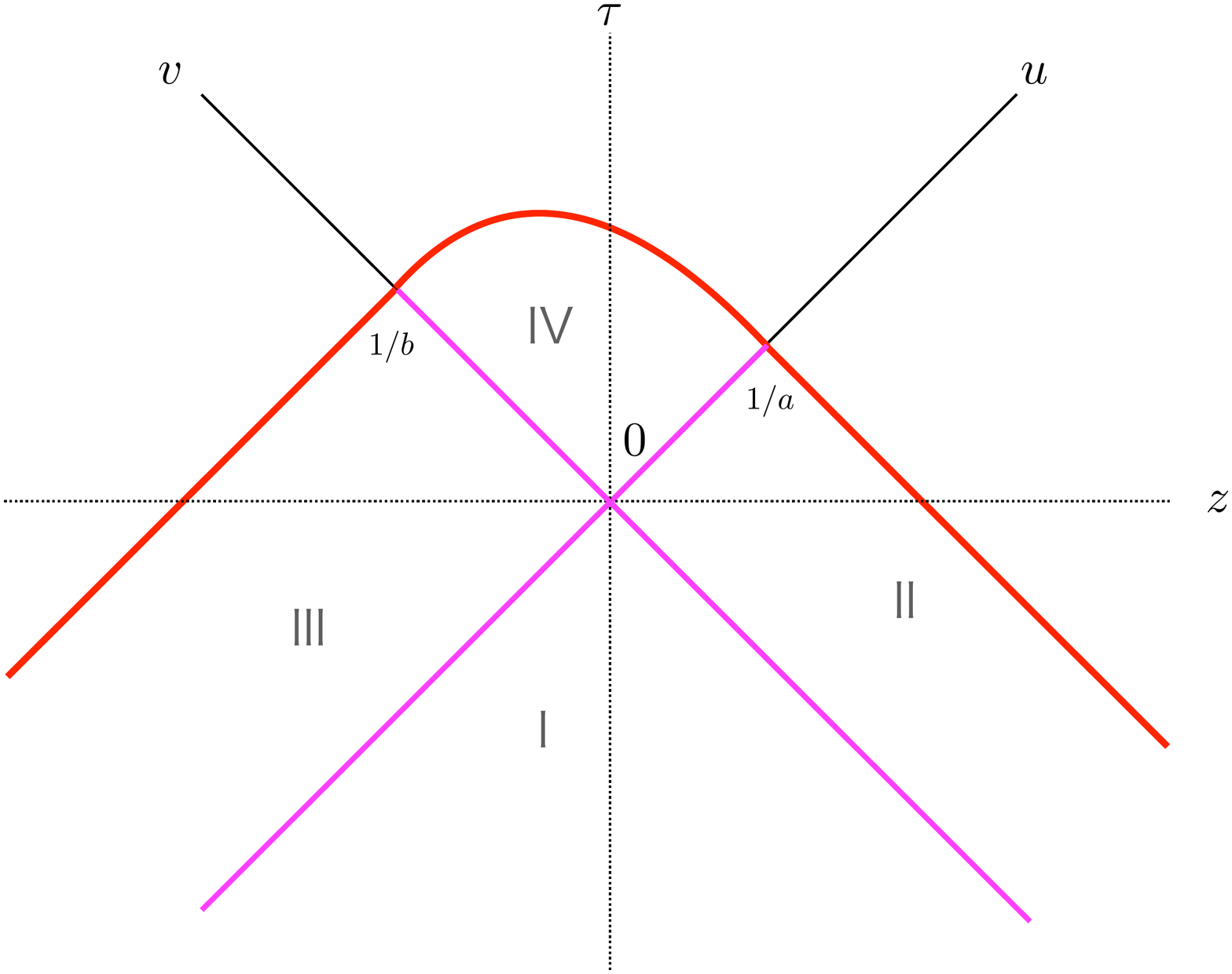}
\end{center}
 \caption{A global spacetime structure in the $u-v$ plane, made of 4 regions: 
 I. Minkowski  ($u<0, v<0$), II. Left moving plane wave  ($u>0, v<0$), III. Right moving plane wave  ($u<0, v>0$),
 IV. Colliding two plane waves  ($u>0, v>0$).
 A red solid line represent singularities at $t^2=0$, while magenta solid lines at $u=0$ and $v=0$ are apparent horizons.
  See the main text for more details.  
 \label{fig:ST1}
  }
 \end{figure}
 Fig.~\ref{fig:ST1} gives a rough sketch of  a spacetime defined by the metric \eqref{eq:metric_rosen}, which exist only at $t^2 \ge 0$ and  is bounded by singularities at $t^2=0$ (a red line in the figure)
as
\beqa
(a u)^{n_1}\theta(u) + (b v)^{n_2} \theta(v) \le 1,
\eeqa  
while apparent horizons appears at $u=0$ with $b v \le 1$ and at $v=0$ with $au\le 1$ (magenta lines in the figure).
 This spacetime is very similar to the one for two colliding impulse waves\cite{Khan:1971vh}
 except that $\delta$ function singularities in the curvature appear at $u=0$ and $v=0$ for two colliding impulse waves\cite{Khan:1971vh},
instead of the apparent horizons mentioned above.
 
\section{Matter energy and generalized matter conserved charge}
\label{sec:charge}
\subsection{Energy momentum tensor at singularities}
A matter energy is evaluated by a volume integral of the EMT at a give $\tau$ as
\[
\int d^3x \sqrt{-g} T^0{}_0 ={1\over 2\kappa} \int d^3x \sqrt{-g} G^0{}_0,
\]
where $\kappa= 4\pi G$ with Newton constant $G$.
While $R^0{}_0$ is zero except singularities, this integral becomes non-zero due to contributions from singularities of $R^0{}_0$.
In order calculate this type of the integral, it is convenient to express $T^\mu{}_\nu$ or $R^\mu{}_\nu$ in terms of delta functions, which
represent contributions from singularities to the integral, as    
in the case of black holes\cite{Balasin:1993fn,Balasin:1993kf,Kawai:1997jx,Aoki:2020prb}.

In order to evaluate a contribution of the EMT in the volume integral, we first regularize singularities by modifying $U,V,M$ in the metric as
\beqa
U_\epsilon &=& - \log T, \quad 
 V_\epsilon = -{k_{12}\over 2} \log T + k_1\log (W+p) +k_2\log (R+q), \nn\\
 M_\epsilon &=& {(4-k_{12}^2)\over 8} \log T +{k_1^2\over 4}\log W + {k_2^2\over 4}\log R +{k_1k_2\over 2} \log(pq+R W),
 \label{eq:reg}
\eeqa
where
\beqa
T(t^2)=\sqrt{t^4 +\epsilon^2}, \quad R^2(r^2)=\sqrt{r^4 +c_r^2\epsilon^2}, \quad  W^2(w^2)=\sqrt{w^4 +c_w^2\epsilon^2},
\label{eq:reg1}
\eeqa
with an infinitesimally small but non-zero real constant $\epsilon$, and real constants $c_r,c_w$.
In the $\epsilon\to 0$ limit, we recover \eqref{eq:sol_IV}.
The regularization of $t^2$ makes all components of Riemann tensor finite at $t^2=0$, so that we can extend the space time to the $t^2<0$ region.
As a result, there may appear other singularities at $r^2=0$ or $w^2=0$ in the region IV in the $t^2<0$ region.
This is a reason why we also have to regularize $r^2$ and $w^2$ by $R^2$ and $W^2$ for the extension of spacetime to  the $t^2<0$ region.
Note that the integral of the EMT tensor in the physical region does not depend on a choice of regularizations after the regularization is removed, as will be seen later. 

With non-zero $\epsilon$, the metric is no more a vacuum solution, giving non-zero EMT as
\beqa
 G^u{}_u &=& G^v{}_v =  - e^{M_\epsilon} f_u g_v {T'' \over T},\\
 G^v{}_u &=& e^{M_\epsilon} f_u^2 \left[ {T''\over T} +\Delta G^v{}_u \right], \quad
 G^u{}_v = e^{M_\epsilon} g_v^2  \left[ {T''\over T} +\Delta G^u{}_v \right], \\
 G^{y_1}{}_{y_1} &=& -e^{M_\epsilon} {(2+k_{12})^2 \over 8}  \left[ {T''\over T} +\Delta G^{y_1}{}_{y_1} \right],\\
 G^{y_2}{}_{y_2} &=& -e^{M_\epsilon} {(2-k_{12})^2 \over 8}  \left[ {T''\over T} +\Delta G^{y_2}{}_{y_2} \right],
 \eeqa
 where extra terms are evaluated for small $\epsilon$ as
 \beqa
 \Delta G^v{}_u &\simeq&  {\epsilon^2 \over 8 T^2} \left[\left( {k_1\over p(p+w)} + {k_2 \over r(r+q)}\right)^2 + O(\epsilon^2)\right],\\
 \Delta G^u{}_v  &\simeq& {\epsilon^2 \over 8 T^2}\left[\left( {k_2\over q(q+r)} + {k_1 \over w(w+p)}\right)^2+ O(\epsilon^2)\right], \\
 \Delta G^{y_1}{}_{y_1} &\simeq& -  {\epsilon^2 \over T^2}\left[ {k_1\over pw(w+p)^2} +{k_2\over rq(r+q)^2}+ O(\epsilon^2)\right],\\
  \Delta G^{y_2}{}_{y_2} &\simeq&   {\epsilon^2 \over T^2} \left[ {k_1\over pw(w+p)^2} +{k_2\over rq(r+q)^2}+ O(\epsilon^2)\right].
 \eeqa
 
 Since
 \beqa
\lim_{\epsilon\to 0} T'' =  \lim_{\epsilon\to 0} {\epsilon^2 \over (t^4+\epsilon^2)^{3/2}} &=&
  \left\{
\begin{array}{lr}
 0, &  t^2\not=0,    \\
 \infty,  &  t^2=0,   \\
\end{array}
\right.
\label{eq:Tdd}
 \eeqa
 and
 \beqa
\int_{0}^\infty dt^2\, T'' = \int_{0}^\infty dt^2\, {\epsilon^2 \over (t^4+\epsilon^2)^{3/2}}
 &=&  \int_0^\infty dy\, {1 \over (y+1)^{3/2}}
 = \int_0^{\pi/2} d\theta \cos\theta = 1,
 \label{eq:Tdd_int}
 \eeqa
 we can write
 \beqa
 \lim_{\epsilon\to 0} T'' =   \lim_{\epsilon\to 0} {\epsilon^2 \over (t^4+\epsilon^2)^{3/2}} &=&  \delta( t^2 ) ,
  \label{eq:T-delta}
 \eeqa
 in the physical region at $t^2\ge 0$.
 Although the explicit form of $T$ in \eqref{eq:reg1} is used to derive \eqref{eq:T-delta},
 one can obtain the same result from a different form of the regularization as long as  $T$ satisfies \eqref{eq:Tdd} and \eqref{eq:Tdd_int}:
 \beqa
 \lim_{\epsilon\to 0} T''(t^2) &=&   \left\{
\begin{array}{lr}
 0, &  t^2\not=0,    \\
 \infty,  &  t^2=0,   \\
\end{array}
\right. \quad
\lim_{\epsilon\to 0}\int_{0}^\infty dt^2\, T''(t^2) = 1 .
 \eeqa

 A non-physical region at $t^2 < 0$ beyond singularities gives
 an extra contribution in this regularization as
 \beqa
 \int_{-\infty}^0 dt^2\, {\epsilon^2 \over (t^4+\epsilon^2)^{3/2}} = 1,
 \eeqa 
which does not affect the integral of  the EMT in the physical region at $t^2\ge 0$.  
An extra contribution depends on how we extend the metric into the $t^2< 0$ region. 
For example, if we take
 \beqa
T(t^2)  &=&
  \left\{
\begin{array}{lr}
 \sqrt{t^4+\epsilon^2}, &  t^2\ge 0,    \\
 \epsilon,  &  t^2 < 0,   \\
\end{array}
\right. ,
 \eeqa
where  $T(0)$ and $T'(0)$ are continuous,
an integral in each region becomes
\beqa
\int^{\infty}_0 dt^2\, T^{\prime\prime}(t^2) =T'(\infty)=1, \quad  \int_{-\infty}^0 dt^2\, T^{\prime\prime}(t^2) = - T'(-\infty) = 0, 
\eeqa
so that an extra contribution from the $t^2<0$ region vanishes.
More generally, if we take
 \beqa
T(t^2)  &=&
  \left\{
\begin{array}{lr}
 \sqrt{t^4+\epsilon^2}, &  t^2\ge 0,    \\
 F_\epsilon(t^2),  &  t^2 < 0,   \\
\end{array}
\right. ,
 \eeqa
 where a function $ F_\epsilon(t^2)$ satisfies
 \beqa
 \lim_{t^2 \to 0^-} F_\epsilon(t^2)=\epsilon, \quad  \lim_{t^2\to 0^-} F_\epsilon^\prime(t^2) =0,
 \eeqa
 the contribution from the non-physical region to the integral becomes
 \beqa
 \int^0_{-\infty} T^{\prime\prime}(t^2) =  \lim_{t^2 \to 0^-}  F_\epsilon^\prime(t^2) - F_\epsilon^\prime(-\infty)   
 = - F_\epsilon^\prime(-\infty) .
 \eeqa
 Thus,  after removing the regularization, 
 the contribution from the non-physical region becomes 
 $ -  \lim_{\epsilon \to 0} F_\epsilon^\prime(-\infty)$,
 which can be arbitrary by a freedom of $F_\epsilon(t^2)$.
Since the physical spacetime at $t^2 \ge 0$ is separated from the non-physical one at $t^2<0$ by singularities at $t^2=0$,
this arbitrariness of the regularization at $t^2<0$ does not affect 
integrals of the EMT in the physical region at $t^2\ge 0$.
 
We can also show that  extra contributions form $\Delta G^a{}_b$ all vanish in the physical region after the $\epsilon \to 0$ limit, since
\beqa
\int_0^\infty dt^2 {\epsilon^2 \over t^4+\epsilon^2} =\epsilon\int _0^\infty {dy \over y^2+1} ={\pi\over 2} \epsilon \to 0.
\eeqa

We then finally obtain 
\beqa
\sqrt{-g}\, G^u{}_u &=& \sqrt{-g}\, G^v{}_v = - f_u g_v \delta(t^2), \quad
\sqrt{-g}\, G^v{}_u = f_u^2 \delta(t^2), \quad
\sqrt{-g}\, G^u{}_v = g_v^2 \delta(t^2), \nn \\
\sqrt{-g}\, G^{y^1}{}_{y^1} &=& - {(k_{12}+2)^2\over 8} f_u g_v \delta(t^2), \quad
\sqrt{-g}\,  G^{y^2}{}_{y^2} = -{(k_{12}-2)^2\over 8} f_u g_v \delta(t^2) 
\label{eq:EMT_final}
\eeqa
in the $\epsilon\to 0$ limit for the original spacetime at $t^2\ge 0$. 

In order to see that the same result \eqref{eq:EMT_final} can be obtained by a simpler regularization \cite{Aoki:2020prb} as
$T(t^2) = t^2 \theta(t^2)$,
 since
 \beqa
 T''(t^2) = 2\delta(t^2) + t^2\delta'(t^2) =\delta(t^2),
 \eeqa 
 where we use $\theta^\prime (x) =\delta(x)$ and $x\delta^{(k)}(x) = - k\delta^{(k-1)}(x)$.
 This agreement again demonstrates that the result \eqref{eq:EMT_final} is universal.
 
 A relation $T^\mu{}_\nu:=\dfrac{1}{2\kappa} G^\mu{}_\nu$ give an expression of the EMT in terms of delta functions.
While singularities are light-like in regions II and III,  singularities in the region IV are space-like, as in the case of the Schwarzschild blackhole.

\
 
\subsection{Energy non-conservation}
 \begin{figure}[hbt]
\begin{center}
\includegraphics[width=0.7\textwidth,clip]{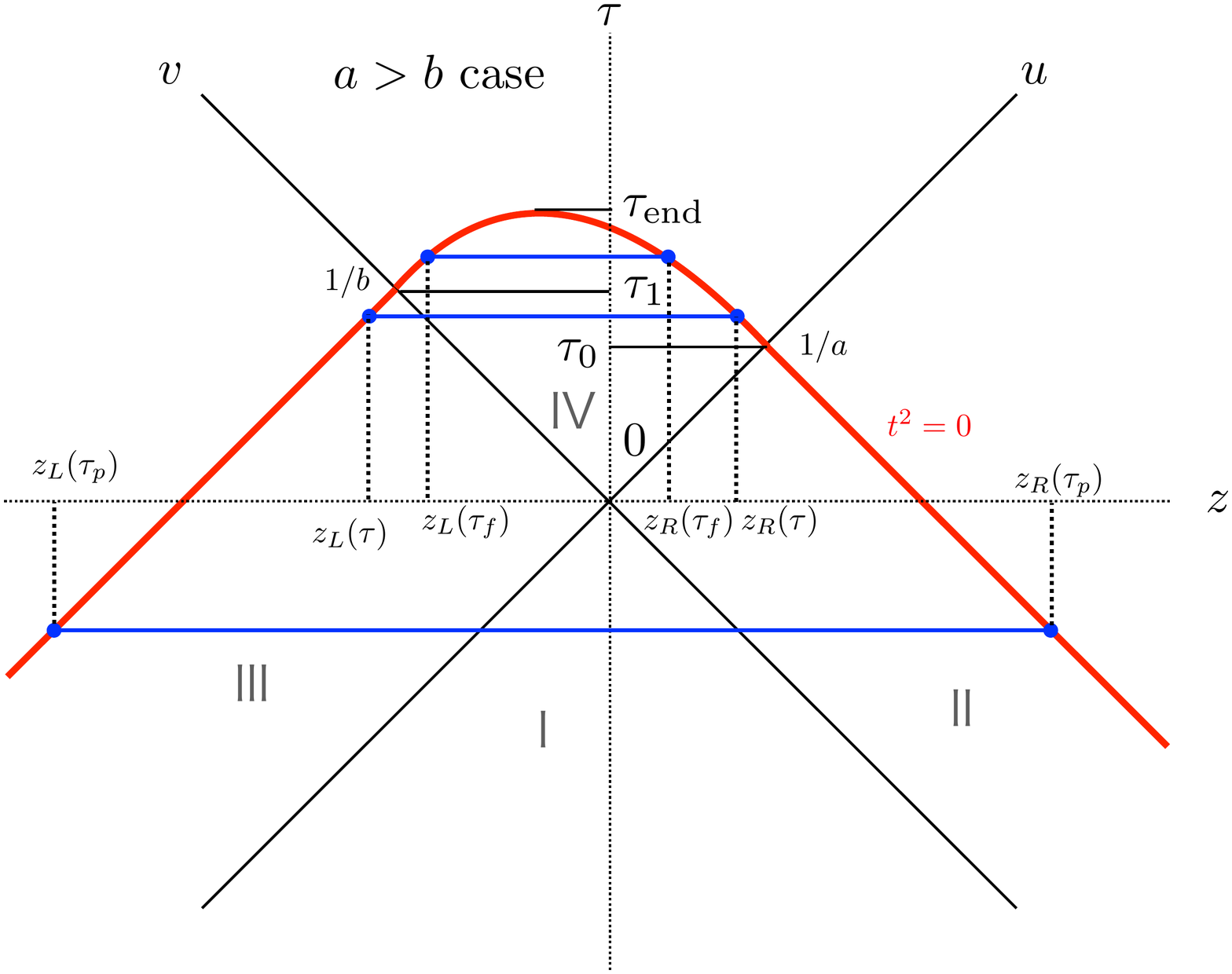}
\end{center}
 \caption{Positions of singularities, $z_L(\tau)$ and $z_R(\tau)$ for several $\tau$, together with $\tau_0,\tau_1,\tau_{\rm end}$,
 defined in the main text.
  \label{fig:ST2}
  }
 \end{figure}
Since the spacetime is uniform in $x$ and $y$ directions, 
we define the matter energy of the system per unit 2-dimensional volume for a given $\tau$
from $T^\tau{}_\mu\xi^\mu$  with $\xi^\mu=\delta^\mu_\tau$\cite{Aoki:2020nzm,Aoki:2020prb}
as
\beqa
E(\tau) &=&{1\over V_2} \int_\tau d^3x\, \sqrt{-g} T^\tau{}_\tau = {1\over 4\kappa}\int_{z_L(\tau)}^{z_R(\tau)} dz\, e^{-M} t^2 
\left( G^u{}_u + G^u{}_v + G^v{}_u +G^v{}_v\right)\nn \\
&=&{1\over 4\kappa}\int_{z_L(\tau)}^{z_R(\tau)} dz\, (f_u-g_v)^2 \delta(t^2), 
\eeqa
where $V_2 :=\int dx dy$ is a 2-dimensional volume, which diverges in infinite 2-dimensional extensions.
The energy defined in the above depends on the choice of the time coordinate $\tau$, as in the case of the flat Minkowski space time, where
the energy also depends on the choice of the time coordinate and it transforms as a vector under Poincare transformation.

In general, $t^2=0$ has two solutions at a given $\tau$ in this spacetime, larger of which is denoted by $z_R(\tau)$ and
smaller by  $z_L(\tau)$, so that  ${d\tau\over dz} < 0$ at $z=z_R(\tau)$ and ${d\tau\over dz} > 0$ at $z=z_L(\tau)$.
Since
\beqa
{d\tau\over dz} = {g_v-f_u\over f_u+ g_v}, \quad f_u+ g_v = -n_1a(au)^{n_1-1} - n_2 b(b v)^{n_2-1}< 0  \quad \mbox{at $t^2=0$},
\eeqa
$g_v-f_u > 0$ at $z=z_R(\tau)$ while $f_u-g_v>0$ at $z=z_L(\tau)$.
At a given $\tau$, the $z$ space exists only in an interval  $ [z_L(\tau),  z_R(\tau)]$.
Positions of $z_{R,L}(\tau)$ depend on a value of $\tau$, as shown in Fig.~\ref{fig:ST2} for $a > b$ and explained below.

\begin{enumerate}
\item $ \tau \le \tau_0$, where $\tau_0 ={1\over \sqrt{2}}\min\left({1\over a}, {1\over b}\right)$.
In this case, $z_R(\tau)={\sqrt{2}\over a} -\tau$ in the region II, while at $z_L(\tau) = - {\sqrt{2}\over b}+\tau$  in the region III.
In Fig.~\ref{fig:ST2}, $z_{R}(\tau_p)$ and $z_{L}(\tau_p)$ belong to this case.

\item $ \tau_0 < \tau \le \tau_1$, where  $\tau_1 ={1\over \sqrt{2}}\max\left({1\over a}, {1\over b}\right)$.
For $a\ge b$, $z_R(\tau)$ is a solution to $f(u)+g(v)=0$ with $f\not= {1\over2}$ and $g\not={1\over 2}$
for a given $\tau$ in the region IV and $z_L(\tau) = -{\sqrt{2}\over b} +\tau$ is a solution of ${1\over 2} + g(v)=0$
in the region III.
In Fig.~\ref{fig:ST2}, $z_{R}(\tau)$ and $z_{L}(\tau)$ belong to this case.
On the other hand, for $a < b$, $z_R(\tau)={\sqrt{2}\over a} -\tau$  is a solution of $f(u)+{1\over 2} =0$ in the region II and $z_L(\tau)$  is a solution to $f(u)+g(v)=0$ for a given $\tau$ in the region IV.

\item $ \tau_1 < \tau \le \tau_{\rm end}$, where  $\tau_{\rm end}$ is a solution to the following equations
\beqa
f(u)+g(v)=0, \quad {d\tau\over d z} = {g_v(v) - f_u(u)\over g_v(v) + f_u(u)} =0 .
\eeqa
Thus $g_v(v)= f_u(u)$.
In a simple case that $n_1=n_2=n$, we have
\beqa
\tau_{\rm end} = {1\over \sqrt{2} ab} \left( a^{n\over n-1} + b^{n\over n-1}\right)^{n-1\over n}.
\eeqa
In this range of $\tau$, $z_{R,L}(\tau)$  are two solutions to $f(u)+g(v)=0$ in the region IV.
In Fig.~\ref{fig:ST2}, $z_{R}(\tau_f)$ and $z_{L}(\tau_f)$ belong to this case.
\end{enumerate}

Using the above property, energy can be evaluated as follows.
At $\tau \le \tau_0$, we have
\beqa
E(\tau) ={\sqrt{2}\over 4\kappa} \left( \left. {f_u^2\over \vert f_u\vert} \right\vert_{z=z_R(\tau)} +  \left. {g_v^2\over \vert g_v\vert}\right\vert_{z=z_L(\tau)}
\right) ={1\over 2\kappa} {( n_1 a+ n_2 b)\over \sqrt{2}}.
\label{eq:E0}
\eeqa
At $\tau_0 < \tau \le \tau_1$, we have
\beqa
E(\tau) &=& 
\left\{
\begin{array}{lc}
 \displaystyle  {1\over 2\kappa}  \left( \left. {g_v-f_u\over \sqrt{2}}\right\vert_{z=z_R(\tau)} + {n_2 b\over \sqrt{2}}\right),
  &   a\ge b , \\
\\
 \displaystyle  {1\over 2\kappa}   \left( {n_1 a\over \sqrt{2}}+\left. {f_u-g_v\over \sqrt{2}}\right\vert_{z=z_L(\tau)}  \right), &  a<b,   \\
\end{array}
\right. .
\label{eq:E1}
\eeqa
Since $g_v-f_u < n_1 a$ for $ua<1$, or $f_u-g_v < n_2 b$ for $vb<1$, $E(\tau)$ in \eqref{eq:E1} 
decreases from \eqref{eq:E0}. Thus the energy is not conserved in this spacetime.\\
At $\tau_1 < \tau \le \tau_{\rm end}$, we finally obtain
\beqa
E(\tau) &=&  {1\over 2\kappa} \left( \left. {g_v-f_u\over \sqrt{2}}\right\vert_{z=z_R(\tau)} +\left. {f_u-g_v\over \sqrt{2}}\right\vert_{z=z_L(\tau)}  \right)
\label{eq:E2},
\eeqa
which is further decreasing, and vanishes at $\tau=\tau_{\rm end}$ when the space in the $z$ direction disappear as $z_R(\tau_{\rm end})=z_L(\tau_{\rm end})$.
Therefore the time $\tau_{\rm end}$ is a moment for an end of the universe, when all matters also should disappear to be consistent with 
$E(\tau_{\rm end})=0$. 

\subsection{Conserved charge}
Even though the matter energy is not conserved, we can construct a non-trivial conserved charge from a conserved current
as $J^\mu := T^\mu{}_\nu \zeta^\nu$, where a vector $\zeta^\nu$ must satisfy $ T^\mu{}_\nu \nabla_\mu \zeta^\nu=0$\cite{Aoki:2020nzm,Aoki:2022gez}.
In this construction, however, there are so many different choices for a direction of the vector $\zeta^\nu$.
In this paper, we propose an unique method to determine $\zeta^\nu$ up to an initial condition.  

We first decompose the EMT as
\beqa
\sqrt{-g} T^\mu{}_\nu &=& -{e^M\over 2\kappa} n^\mu n_\nu \delta(t^2) +\cdots,
\eeqa
where ellipses represents $x,y$ components, which we do not consider in this section, and vectors $n^\mu$ and $n_\nu$ are given by
\beqa
n^\mu = (g_v, -f_u,0,0), \quad n_\nu := g_{\nu\mu} n^\mu = e^{-M}(f_u,-g_v,0,0)
\eeqa 
in the $(u,v,x,y)$ coordinate.

Our new proposal is to take $\zeta^\mu \propto n^\mu$, which is determined by the EMT, thus is coordinate independent.
Explicitly, we take
\beqa
\zeta^\mu &=& -\kappa {\beta(u,v)\over f_u g_v} n^\mu
\eeqa
in the region IV, where $ f_u g_v\not=0$.
Then the conserved current density becomes
\beqa
{\cal  J}^\mu := \sqrt{-g} T^\mu{}_\nu \zeta^\nu = (g_v, -f_u,0,0) \beta(u,v)\delta(t^2),
\eeqa
which can be used even in regions II and III.

Since we require $\sqrt{-g}\nabla_\mu T^\mu{}_\nu \zeta^\nu = \partial_\mu {\cal J}^\mu =0$, $\beta$ must satisfies
\beqa
g_v \partial_u\beta -f_u \partial_v \beta =0,
\eeqa
whose general solution is given by
\beqa
\beta(u,v) = \beta_0( f+g),
\eeqa
where $\beta_0(t^2)$ is an arbitrary differentiable function.

The corresponding conserved charge per $V_2$ is defined by
\beqa
S(\tau) &:=& {1\over V_2} \int_\tau d^3 x\,  \sqrt{-g} T^\tau{}_\nu \zeta^\nu =
{1\over \sqrt{2}}  \int_\tau dz\, ({\cal J}^u + {\cal J}^v)
= {1\over \sqrt{2}}  \int_\tau dz (g_v - f_u) \beta_0( f+g) \delta(f+g).\nn \\
\eeqa
The conservation of $S(\tau)$ is derived from an integral of $\partial_\mu {\cal J}^\mu =0$ over a space-time region
in the $z-\tau$ plane surrounded by 4 boundaries,
$[( z_L(\tau_f),\tau_f) , (z_R(\tau_f),\tau_f)]$,  $[(z_R(\tau_f),\tau_f),  (z_R(\tau_p),\tau_p)]$,  $[(z_R(\tau_p),\tau_p), (z_L(\tau_p),\tau_p)]$, and 
$[(z_L(\tau_p),\tau_p) , (z_L(\tau_f),\tau_f)]$, as shown in Fig.~\ref{fig:ST2}.
Therefore, we need to check that boundary contributions on $[(z_R(\tau_f),\tau_f) , (z_R(\tau_p),\tau_p)]$ and $[(z_L(\tau_p),\tau_p), (z_L(\tau_f),\tau_f)]$  vanish
to establish the conservation of $S(\tau)$. Since the (unnormalized)  normal vector to the singularity surface is given by
$n_\mu = \partial_\mu t^2= (f_u,g_v, 0,0)$, integrands in boundary contributions along the singularity surface always vanish as
\beqa
n_\mu {\cal J}^\mu  \propto  (f_u, g_v, 0,0) ^T \cdot (g_v, -f_u,0,0)  = 0, 
\eeqa 
so that $S(\tau_f) = S(\tau_p)$ for arbitrary $\tau_f,\tau_p$.

The charge $S(\tau)$ at all $\tau \le \tau_{\rm end}$  can be explicitly calculated as
\beqa
S(\tau) &=& \beta_0(0)\left[ \left. {g_v-f_u\over \vert g_v-f_u\vert}\right\vert_{z=z_R(\tau)}+ \left. {g_v-f_u\over \vert g_v-f_u\vert}\right\vert_{z=z_L(\tau)}  \right] =\beta_0(0) [ 1 - 1] = 0,
\label{eq:S_end}
\eeqa 
which vanishes due to a cancelation between a contribution at $z_R(\tau)$  and a  contribution at $z_L(\tau)$,
except $\tau=\tau_{\rm end}$ where a contribution identically vanishes at $z=z_R(\tau_{\rm end})=z_L(\tau_{\rm end})$.
The charge $S(\tau)$ is indeed conserved.

We thus conclude that there exists a conserved charge $S(\tau)$ in the spacetime given by \eqref{eq:metric_rosen}, whose value is zero by the cancellation of two contributions.
A fact that $S(\tau)=0$ in this spacetime seems natural, since the spacetime ceases to exist at $\tau=\tau_{\rm end}$, when all matters should disappear,
so that $S(\tau_{\rm end})=E(\tau_{\rm end})=0$. If $S(\tau)$ is conserved unlike energy $E(\tau)$, $S(\tau)=0$ must hold for all $\tau$.

In \cite{Aoki:2020nzm,Aoki:2022gez}, analysis for some particular spacetimes with matters described by a perfect fluid  indicated that a generic matter conserved charge $S(\tau)$ in general relativity may be interpreted as a matter  {\it entropy}.
Since $S(\tau)$ can be locally negative in the metric \eqref{eq:metric_rosen}, however, we may need to reconsider a physical interpretation of the matter conserved charge $S(\tau)$ once more. 
We may call $S(\tau)$ a gravitational charge, which is  more general and
becomes the matter entropy only for special cases. 
The gravitational charge may be locally negative
in the special spacetime described by \eqref{eq:metric_rosen}, which ceases to exist at $\tau=\tau_{\rm end}$ and whose singularities in the region IV are superluminal.
Further investigations will be required to distinguish one possibility from the other.

\section{Construction of new spacetimes} 
\label{sec:New}
Motivated by the geometry of colliding plane gravitational waves, we construct other types of the plane wave(s) in this section.
Since $n_1$ and $n_2$ are even integers, 
$f(u)=f_0(u)$ and $g(v)=g_0(v)$ in  \eqref{eq:sol_IV} and \eqref{eq:pqrw} satisfy  the vacuum Einstein equation 
in all regions, where
\beqa
f_0(u) &=& {1\over 2} -(au)^{n_1}=f_0(-u), \quad g_0(v)={1\over 2} -(b v)^{n_2}=g_0(-v).
\eeqa
We can construct new solutions by putting $\theta$ functions to appropriate places, in order to satisfy boundary conditions at $u=0$ and $v=0$.

\subsection{A scattering of the plane gravitational wave} 
 \begin{figure}[htb]
\begin{center}
\includegraphics[width=0.49\textwidth,clip]{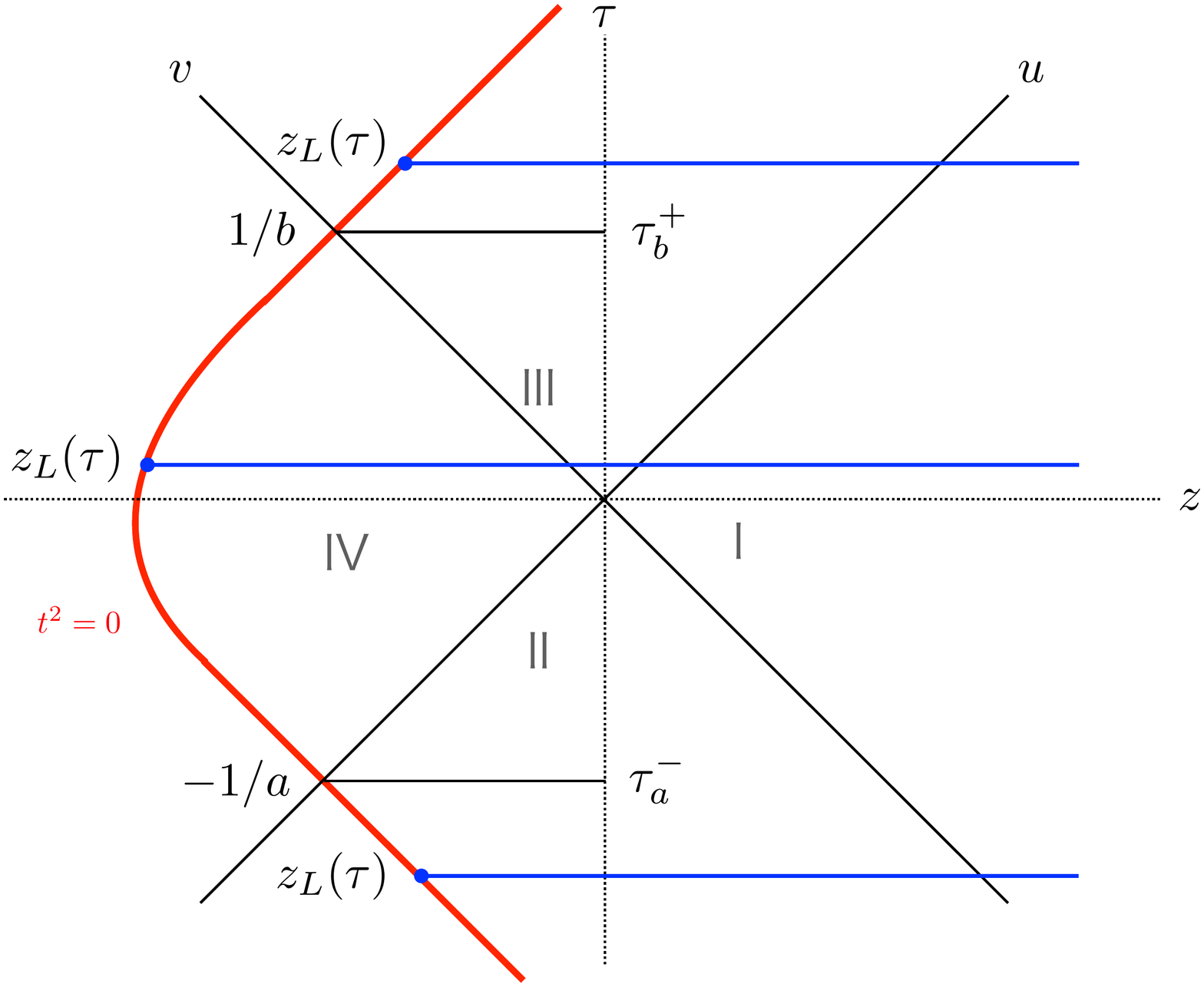}
\includegraphics[width=0.49\textwidth,clip]{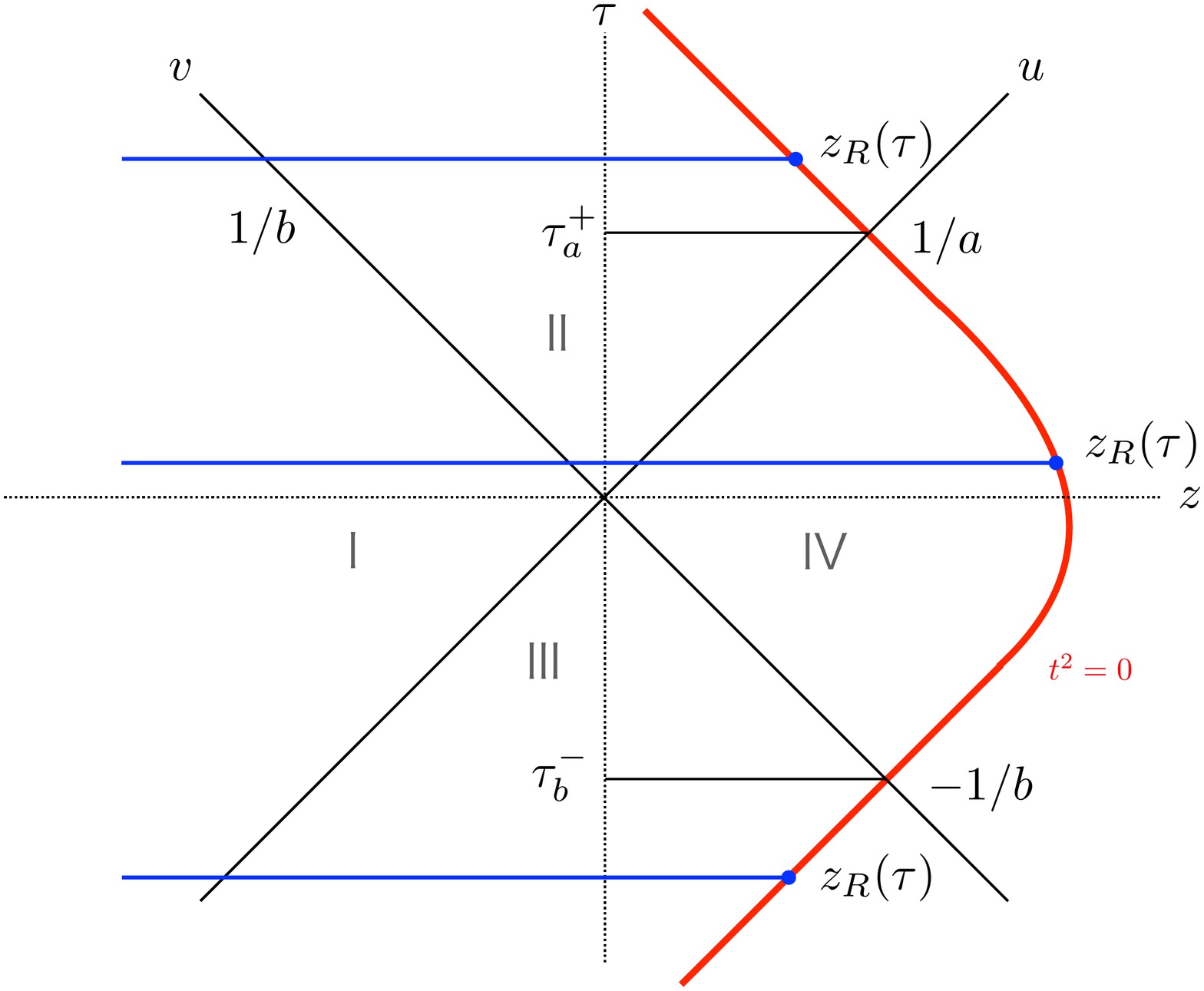}
\end{center}
 \caption{A global spacetime structure for a scattered plane gravitational wave.
 (Left) The left-moving gravitational wave is scattered to be the right-moving one.
 (Right) From the right-moving to the left-moving. 
 \label{fig:ST3}
  }
 \end{figure}
 The first example is a scattering of the plane gravitational wave.
 
The left-moving gravitational wave in the region II is scattered in the region IV, and then it appears as
the right-moving gravitational wave in the region III, as shown in Fig.~\ref{fig:ST3} (Left),
while the right-moving gravitational wave in the region III, scattered in the region IV, goes into the region II as the left-moving wave, as shown 
in Fig.~\ref{fig:ST3} (Right).  Solutions for both cases are given by \eqref{eq:sol_IV} with
\beqa
f(u) =f_-(u):={1\over 2} -(-au)^{n_1}\theta(-u), \quad g(v) =g_+(v):= {1\over 2} -(bv)^{n_2}\theta(v)
\eeqa
for the left-to-right scattering, and  
\beqa
f(u) =f_+(u):={1\over 2} -(au)^{n_1}\theta(u), \quad g(v)=g_-(v): = {1\over 2} -(-bv)^{n_2}\theta(-v)
\eeqa
for the right-to-left scattering. Note that singularities at $t^2=0$ (red line) are either time-like in the region IV or light-like in regions II and III.

Let us calculate the matter energy using the definition as
\beqa
E(\tau) =  {1\over 4\kappa} \int_{\tau} dz (f_u-g_v)^2 \delta(t^2).
\eeqa

The matter energy for  the left-to-right scattering in Fig.~\ref{fig:ST3} (Left) becomes
\beqa
E(\tau) = {1\over 4\kappa} \int_{z_L(\tau)}^{+\infty} dz f_u^2\delta( q^2) =
{\sqrt{2} \over 4\kappa}\left. {f_u^2\over \vert f_u\vert}\right\vert_{z=z_L(\tau)}
=   {1\over 2\kappa}{ n_1 a \over \sqrt{2}}
\eeqa
at  $\tau\le \tau_a^- := -\dfrac{1}{\sqrt{2} a}$ (the region II), and
\beqa
E(\tau)= {1\over 2 \kappa}  {n_2 b\over \sqrt{2}}
\eeqa
at $\tau_b^+ :=\dfrac{1}{\sqrt{2} b} \le \tau$ (the region III). 
In the region IV at $\tau_a^- < \tau < \tau_b^+$, we have
\beqa
E(\tau)= {1\over 2 \kappa} \left. {f_u - g_v\over \sqrt{2}}\right\vert_{z=z_L(\tau)} .
\eeqa 
Therefore the matter energy $E(\tau)$ is constant in the region II as well as the region III, while
$E(\tau)$ is decreasing or increasing in the region IV if $ n_1a> n_2b$ or $ n_1a <n_2 b$, respectively.

Similarly, the matter energy for  the right-to-left scattering in Fig.~\ref{fig:ST3} (Right) is given by 
\beqa
E(\tau) &=&\displaystyle {1\over 2\sqrt{2}\kappa} \times \left\{
\begin{array}{lr}
  n_2 b,    &  \tau\le \tau_b^- ,  \\
  \\
(g_v-f_u)\vert_{z=z_R(\tau) },   & \tau_b^- < \tau < \tau_a^+,   \\
 \\
 n_1 a ,&  \tau_a^+ \le \tau,  \\   
\end{array}
\right.
\eeqa
where $\tau_b^- := -\dfrac{1}{\sqrt{2} b}$ and $ \tau_a^+ := \dfrac{1}{\sqrt{2} a}$.
Again the matter energy $E(\tau)$ is constant in the regions III and II, while
$E(\tau)$ is decreasing or increasing in the region IV for  $n_1a < n_2b$ or $ n_1a > n_2 b$, respectively.
 
\vskip 0.5cm 

We next consider the matter conserved charged, given by
\beqa
S(\tau) ={\beta_0(0)\over \sqrt{2}} \int dz\, (g_v - f_u) \delta(t^2) .
\eeqa
In the case of the left-to-right scattering, we obtain
\beqa
S(\tau) = \beta_0^{LR}{ g_v - f_u\over \vert f_u -g_v\vert} = -  \beta_0^{LR}, \quad \beta^{RL}_0:=\beta_0(0),
\label{eq:S_LR}
\eeqa
while we have
\beqa
S(\tau) =  \beta_0^{RL}, \quad \beta_0^{RL}:=\beta_0(0),
\label{eq:S_RL}
\eeqa
for the right-to-left scattering, where $\beta_0^{LR}$ and $\beta_0^{RL}$ are arbitrary constants given by $\beta_0(0)$. 
Therefore the matter conserved charge is non-zero and conserved for these scatterings. 

\subsection{Pair creation of gravitational waves}
 \begin{figure}[htb]
\begin{center}
\includegraphics[width=0.7\textwidth,clip]{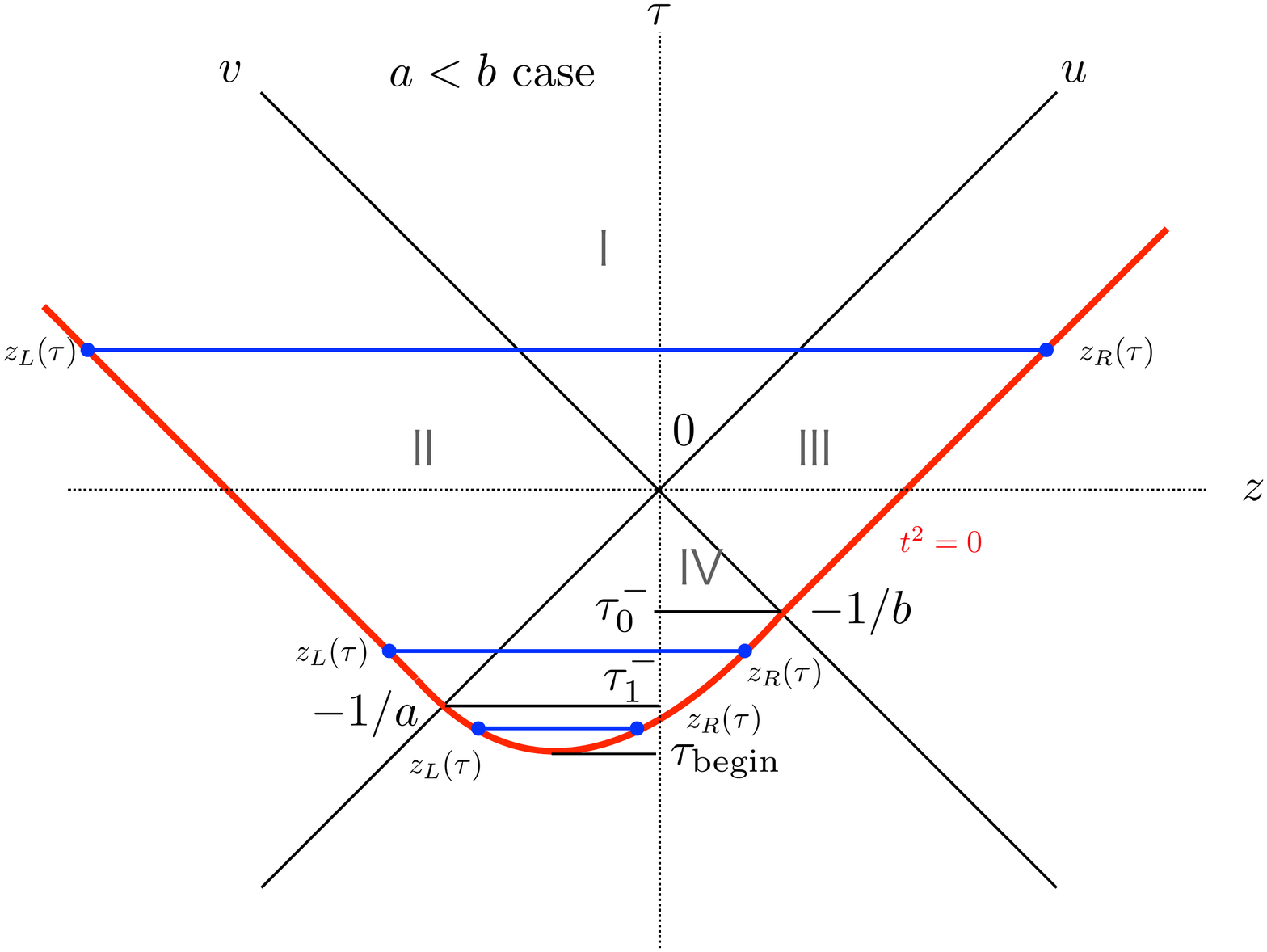}
\end{center}
 \caption{A global spacetime structure for a pair creation of plane gravitational waves.
  \label{fig:ST5}
  }
 \end{figure}
As a time reversal process to the colliding plane gravitational waves, we consider a creation of left and right moving plane gravitational waves,
as illustrated in Fig.~\ref{fig:ST5}, where a pair creation occurs at $\tau=\tau_{\rm begin}$ in the region IV.

With functions $f(u)=f_-(u)$ and $g(v)=g_-(v)$,
$\tau_{\rm begin}$ is determined from $u$ and $v$ which satisfy
\beqa
t^2=f(u) + g(v)=0, \quad f_u(u) = g_v(v).
\eeqa
For a special case that $n_1=n_2=n$, we obtain
\beqa
\tau_{\rm begin} = -{1\over \sqrt{2} ab} \left( a^{n\over n-1} + b^{n\over n-1}\right)^{n-1\over n}.
\eeqa

At $\tau_{\rm begin} \le \tau < \tau_1^- := \min\left(\tau_a^-, \tau_b^- \right)$,
the matter energy at $\tau$ in this range is given by
\beqa
E (\tau)&=& {1\over 2\sqrt{2}\kappa}\left[ (f_u-g_v)\vert_{z=z_L(\tau)} +  (g_v-f_u)\vert_{z=z_R(\tau)}\right],
\eeqa
which increases from $E(\tau_{\rm begin})=0$ as $\tau$ increases,
where $z_L(\tau)$ and  $z_R(\tau)$ are smaller and larger solutions to $t^2 = 0$ in the region IV, respectively.
 
At $\tau_1^- \le \tau  < \tau_0^- :=  \max\left( \tau_a^-, \tau_b^- \right)$, we have
\beqa
E(\tau) &=&  {1\over 2\sqrt{2}\kappa} \times \left\{
\begin{array}{cc}
\left[ n_1 a +  (g_v-f_u)\vert_{z=z_R(\tau)} \right], & a \le b \\
\\
 \left[(f_u-g_v)\vert_{z=z_L(\tau)} + n_2 b\right],& a > b \\
\end{array}
\right. ,
\eeqa
which is still increasing, since $ (f_u-g_v)\vert_{z=z_L(\tau)} \le n_1 a$ and $(g_v-f_u)\vert_{z=z_R(\tau)} \le n_2 b$.

Finally the matter energy becomes largest at $\tau=\tau_0^-$, and stays constant at $\tau \ge \tau_0^- $ as
\beqa
E(\tau) &=&  {1\over 2\sqrt{2}\kappa}\left[ n_1 a +  n_2 b\right].
\eeqa

\vskip 0.5cm 
As in the case of the colliding plane gravitational wave, the matter conserved charge $S(\tau)$ at all $\tau \ge \tau_{\rm begin}$ vanishes as
\beqa
S(\tau) &=&  {\beta_0(0) \over \sqrt{2}}\int_{z_L(\tau)}^{z_R(\tau)} dz\, (g_v - f_u) \delta(f+g)
=   \beta_0(0) \left[ \left. {g_v-f_u \over \vert f_u-g_v\vert}\right\vert_{z=z_L(\tau)} +  \left. {g_v-f_u \over \vert f_u-g_v\vert}\right\vert_{z=z_R(\tau)}\right] \nn \\
&=&  \beta_0(0)\left[ -1 + 1\right] =0.
\label{eq:S_begin}
\eeqa 

\subsection{Minkowski vacuum bottle}
 \begin{figure}[htb]
\begin{center}
\includegraphics[width=0.9\textwidth,clip]{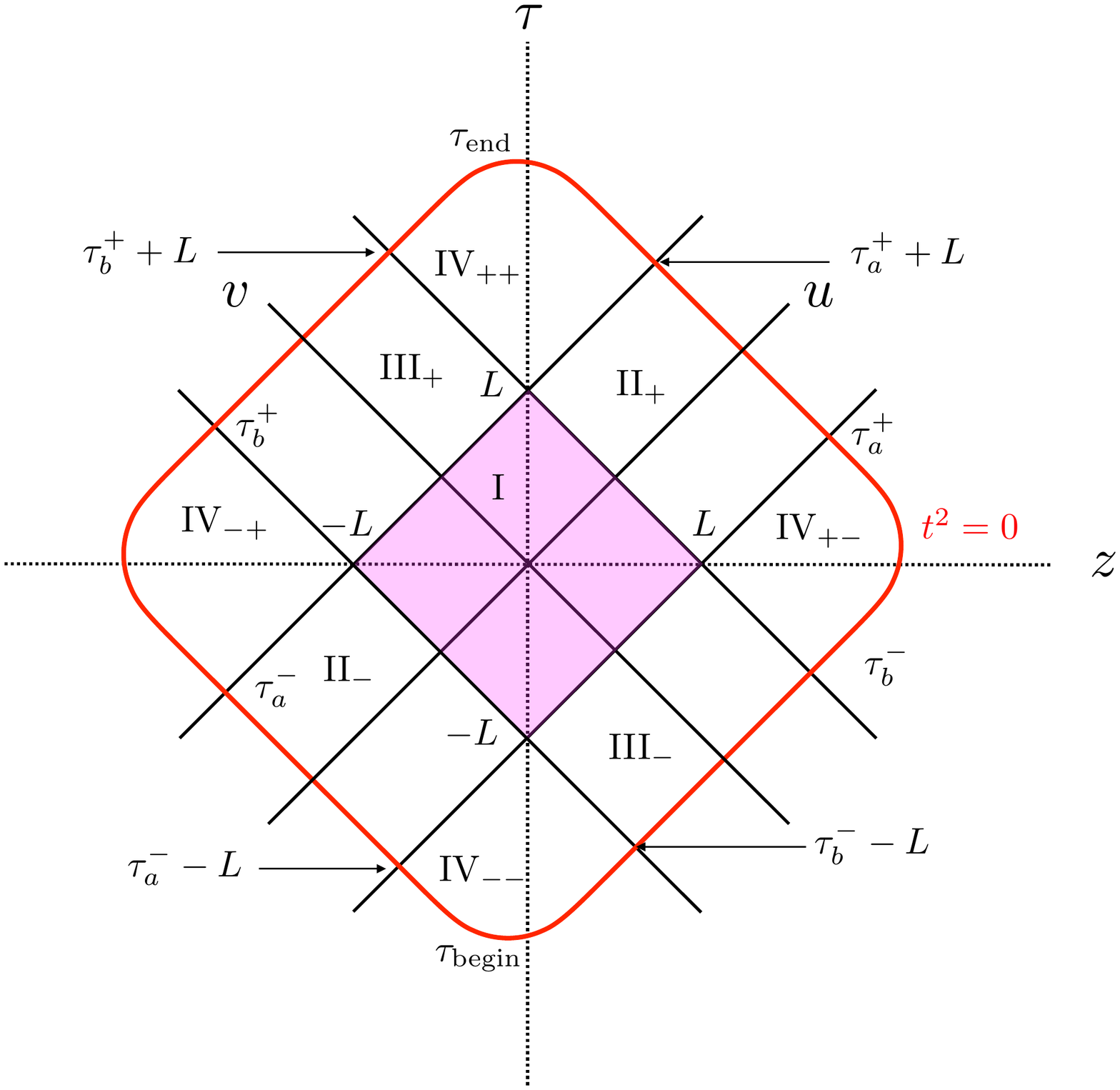}
\end{center}
 \caption{A global spacetime structure for a Minkowski vacuum bottle. Some time coordinates such as $\tau_a^\pm$  and $\tau_b^\pm$ are explicitly written in the figure.
  \label{fig:ST6}
  }
 \end{figure}
 We can assemble all four spacetimes: two gravitational waves are created at $\tau=\tau_{\rm begin}$, then the left-moving wave is scattered to the right-moving and vice versa, and two waves collides and finally annihilate at $\tau=\tau_{\rm end}$, as illustrated in Fig.~\ref{fig:ST6},
 where there are 9 regions denoted by I, II$_{\pm}$, III$_{\pm}$, IV$_{\pm\pm}$ and IV$_{\pm\mp}$. 
 
 To specify these regions, we need 4 functions, $f^\pm(u)$ and $g^\pm(v)$, defined by
 \beqa
 f^\pm(u) = h( u_\pm,a_\pm,n_{1}^\pm), \quad  g^\pm(v) = h( v_\pm,b^\pm,n_{2}^\pm),
 \eeqa
 where 
 \beqa
 h(x, a, n) ={1\over 2} -(ax)^n, \quad u_{\pm} =\pm { \tau \mp L +z\over \sqrt{2}},   \ v_\pm = \pm { \tau \mp L - z\over \sqrt{2}},
 \eeqa
 and $2L$ is a diagonal length of the squire region I in the middle, whose center is $(z,\tau) = (0,0)$. 
 A function $f^\alpha$ corresponds to a region II$_\alpha$, while $g^\beta$ corresponds to a region III$_\beta$.
 On the other hand, a pair of function $(f^\alpha, g^\beta)$ are used to specify a region IV$_{\alpha\beta}$.

During the time evolution, the matter energy $E(\tau)$ first increase from $E(\tau_{\rm begin}) =0$ ,  then decreases and
vanishes at $\tau=\tau_{\rm end}$, where $\tau_{\rm begin}$ and $\tau_{\rm end}$ satisfies 
$f^-_u(u)=g^-_v(v)$ with $f^-(u)+g^-(v)=0$ and $f^+_u(u)=g^+_v(v)$ with $f^+(u)+g^+(v)=0$, respectively.
Note that energy sometimes stays constant for a while during this process.

Explicitly we have $E(\tau) = E_{LR}(\tau) + E_{RL}(\tau)$,
where $E_{LR}(\tau)$ and $E_{RL}(\tau)$ are energies of the left-to-right moving wave and the right-to-left moving one, respectively, which are given by
\beqa
E_{LR}(\tau) &=&  {1\over 2\sqrt{2}\kappa} \times \left\{
\begin{array}{cc}
 (f^-_u-g^-_v)\vert_{z=z^{--}_L(\tau)}, & \tau_{\rm begin}  \le \tau < \tau_a^- -L\\
 \\
n_{1}^- a_-,  &  \tau_a^- -L \le \tau <  \tau_a^-, \\
\\
(f^-_u -g^+_v)\vert_{z=z^{-+}_L(\tau)}, & \tau_a^- \le \tau < \tau_b^+, \\ 
\\
n_2^+ b_+, & \tau_b^+ \le \tau < \tau_b^+ + L, \\
\\
(f^+_u -g^+_v)\vert_{z=z^{++}_L(\tau)}, & \tau_b^+ + L\le \tau \le \tau_{\rm end}, \\
\end{array}
\right. ,
\eeqa
\beqa
E_{RL}(\tau) &=&  {1\over 2\sqrt{2}\kappa} \times \left\{
\begin{array}{cc}
- (f^-_u-g^-_v)\vert_{z=z^{--}_R(\tau)}, & \tau_{\rm begin}  \le \tau < \tau_b^- -L\\
 \\
n_{2}^- b_-,  &  \tau_b^- -L \le \tau <  \tau_b^-, \\
\\
-(f^+_u -g^-_v)\vert_{z=z^{+-}_R(\tau)}, & \tau_b^- \le \tau < \tau_a^+, \\ 
\\
n_1^+ a_+, & \tau_a^+ \le \tau < \tau_a^+ + L, \\
\\
-(f^+_u -g^+_v)\vert_{z=z^{++}_R(\tau)}, & \tau_a^+ + L\le \tau \le \tau_{\rm end}, \\
\end{array}
\right. .
\eeqa
Here $z^{\alpha\beta}_L(\tau)$ and $z^{\alpha\beta}_R(\tau)$  are smaller and larger solutions to $f^\alpha(u) + g^\beta(v) = 0$ at a given $\tau$ for $\alpha,\beta$ take $\pm$. See some time coordinates in  Fig.~\ref{fig:ST6}.

The matter conserved charge $S(\tau)$ is always zero by a cancellation between the left-to-right moving wave and the right-to-left moving wave,
as seen from  \eqref{eq:S_end}, \eqref{eq:S_begin}, and \eqref{eq:S_LR} $+$  \eqref{eq:S_RL}
with $\beta_0^{LR} = \beta_0^{RL} =\beta_0(0)$. 

It is interesting to see that the Minkowski vacuum  appears in the center of this spacetime ( the region I in magenta) surrounded by two moving plane gravitational waves with matters at singularities. Thus we call this spacetime a "Minkowski vacuum bottle".
The region I is separated from others by apparent horizons, and begins at $\tau=-L$, expands until  $\tau=0$, then contracts and finally disappears at $\tau=L$.  
  
\section{Conclusions and Discussions}  
\label{sec:Concl}
In this paper, we have analyzed a model of colliding plain gravitational waves, proposed by Szekeres~\cite{Szekeres:1972uu}.
A structure of singularities in the spacetime is determined, and  then
contributions from the energy momentum tensor (EMT) at these singularities are  determined through the Einstein equation.  
We have evaluated the total energy of the matter  as a volume integral of the EMT\cite{Aoki:2020prb, Aoki:2020nzm}, 
which is conserved before  the collision of two plane gravitational waves but decreases during the collision and becomes zero at the end of the collision.
Thus the model of colliding plain gravitational waves can be regarded as a spacetime describing a pair annihilation of plain gravitational waves.
We have also evaluated the gravitational charge as the generalized matter conserved charge proposed in Ref.~\cite{Aoki:2020nzm}.
The gravitational charge is indeed conserved but is zero due to a cancellation of contributions between two plain gravitational waves.
The vanishing conserved charge seems natural since nothing remains after a pair annihilation of plain gravitational waves.
While the gravitational charge can be interpreted as  the entropy  in Ref.~\cite{Aoki:2020nzm} for special cases,
our result in this paper that it can become locally negative may suggest that  the gravitational charge is more general than the entropy.
We leave this problem to future studies. 
It is also interesting to apply the analysis in this paper to other models of colliding gravitational waves  
such as \cite{Alcubierre:1999ex,Pretorius:2018lfb}, for example.

By modifying the spacetime for a pair annihilation of plain gravitational waves, we construct two types of a scattering plane gravitational wave as well as
a pair creation of plain gravitational waves. Combining all, we also create a Minkowski vacuum bottle, a Minkowski spacetime surrounded by two moving plane gravitational waves with singularities. The total matter energy as well as the conserved gravitational charge are calculated in each case.
As expected, while the matter energy is not conserved, the gravitational charge is indeed conserved.

According to the proposal in Ref.~\cite{Aoki:2020prb, Aoki:2020nzm},  an answer to the question in the beginning of the introduction 
that "Is a total energy always conserved in general relativity ? " may be answered as follow.
In general relativity, 
the total energy of the matter can be defined but is not conserved in general. 
However there always exists a conserved gravitational charged as a matter conserved charge.
In this paper we have shown that these statements hold for  the colliding plain gravitational waves and their variants.
The above answer, however, seems unsatisfactory, since one may hope that the total energy including both matter energy and gravitational energy is conserved in general relativity.  
Unfortunately, it is not easy to define the gravitational energy and thus the conserved total energy in general relativity.
One may say that the conserved current of the Noether's 2nd theorem can be used to define the total energy in general relativity.
This, however, may not be the final answer since the conservation via the Noether's 2nd theorem is not dynamical\cite{Noether:1918zz,Aoki:2022gez,Deriglazov:2017biu}.
As an evidence to support this statement, 
we show in appendix~\ref{app:Komar} that
the generalized Komar integral\cite{Komar:1958wp}, which is a generic conserved charge from the Noether's 2nd theorem and is regarded as a  representative for the quasi-local energy, is not conserved for the colliding plane gravitational waves.
This suggests that the generalized Komar integral is physically an improper definition of ``energy", as has been pointed out before\cite{Misner:1963zz}.
To find a conserved total energy including contributions from gravitational fields will be the next important task in our future studies,
even though such a concept may not exist in general relativity.   

\section*{Acknowledgment}
The author would like to thank Mr. Takumi Hayashi , Drs. Ryo Namba, Naritaka Oshita and Daisuke Yoshida for useful discussions and valuable comments, and give his special thanks to Mr. Takumi Hayashi for his contributions, who provided the result in eq.~\eqref{eq:RR} by Mathematica  before the author's analytic calculation.
This work is supported in part by the Grant-in-Aid of the Japanese Ministry of Education, Sciences and Technology, Sports and Culture (MEXT) for Scientific Research (Nos.~JP22H00129).

\appendix
\section{Generalized Komar integral for colliding plane gravitational waves}
\label{app:Komar}
In this appendix, we evaluate a Komar-type integral\cite{Komar:1958wp} for the colliding plane gravitational waves,
since it, together with its variants, covers large varieties of quasi-local definitions of ``energy" in general relativity.

The Komar current\footnote{The Komar energy is originally defined  by $K^\mu[\xi]$ in the case that $\xi$ is a time-like Killing vector.
 Since $K^\mu[\xi]$ is always conserved, we here define the Komar energy for an arbitrary vector $\xi^\mu$, which we call the generalized Komar integral.}
 is given by
\beqa
K^\mu[\xi] &:=& {1\over 2\kappa} \nabla_\nu K^{\mu\nu}[\xi], \quad K^{\mu\nu}[\xi]:= \nabla^{[\mu}\xi^{\nu]},
\eeqa
 whose covariant divergence identically vanishes as $\nabla_\mu K^\mu[\xi]=0 $ for an arbitrary vector $\xi^\mu$ without using equations of motion, 
 as a consequence of Noether's 2nd theorem\cite{Aoki:2022gez}.
 Therefore an identity, 
 \beqa
 0 &=& \int d^4x\, \sqrt{-g} \nabla_\mu K^\mu[\xi] = \int_{\partial V} [d^3x]_\mu K^\mu[\xi],
 \eeqa
holds for an arbitrary space-time region $V$ with a boundary $\partial V$.
Taking a region in the $\tau-z$ plane surrounded by 4 boundaries, $[z_L(\tau_p), z_R(\tau_p)]$, $ [z_R(\tau_p), z_R(\tau_f)]$, $[z_R(\tau_f), z_L(\tau_f)]$ and $[z_L(\tau_f), z_L(\tau_p)]$ (See an example in Fig.~\ref{fig:ST2}), we obtain a relation that
\beqa
E_K(\tau_f) - E_K(\tau_p) &=& S_R(\tau_f,\tau_p) - S_L(\tau_f,\tau_p),
\eeqa 
where
\beqa
E_K(\tau) &:=& {1\over V_2}\int_{z_L(\tau)}^{z_R(\tau)} [d^3 x]_\mu K^\mu[\xi], \
S_{H}(\tau_f,\tau_p) :=  {1\over V_2} \int_{z_{H}(\tau_p)}^{z_{H}(\tau_f)}  [d^3 x]_\mu K^\mu[\xi], \ H=R,L.~~~~
\eeqa
Therefore, the generalized Komar integral $E_K(\tau)$ is $\tau$-independent ({\it i.e.} conserved), if $S_R(\tau_f,\tau_p) -  S_L(\tau_f,\tau_p)=0$ for $^\forall\tau_f, ^\forall\tau_p$.

Using the Stokes theorem and performing trivial $x,y$ integrals,  we write
\beqa
E_K(\tau) &=& \left. {1\over 2\kappa} \sqrt{-g} K^{vu}[\xi]\right\vert_{z_L(\tau)}^{z_R(\tau)}, \quad
S_H(\tau_f,\tau_p) =  \left. {1\over 2\kappa} \sqrt{-g} K^{vu}[\xi]\right\vert_{z_H(\tau_p)}^{z_H(\tau_f)}, 
\eeqa
where
\beqa
K^{vu}[\xi] &=& g^{vu}(\partial_u \xi^u +\Gamma^u_{uu}\xi^u)- g^{uv}(\partial_v \xi^v +\Gamma^v_{vv}\xi^v).
\eeqa
In the following analysis, we take $\xi^u=h_1(v)$ and $\xi^v =h_2(u)$, so that
\beqa
\sqrt{-g}\,  K^{vu}[\xi] &=&  \left[ M_u h_1(v) - M_v h_2(u)\right],
\eeqa
where
\beqa
M_u &=& f_u\left[1-{k_{12}^2\over 4} +O(t^2) \right], \quad
M_v = f_v\left[1-{k_{12}^2\over 4} +O(t^2) \right],
\eeqa
at $t^2\simeq 0$.  We thus obtain 
\beqa
E_K(\tau) &=&\left. {1\over 4\kappa} \left(1-{k_{12}^2\over 4}\right) (f_u h_1 -g_v h_2 )\right\vert_{z_L(\tau)}^{z_R(\tau)},\\
S_H(\tau_f,\tau_p) &=&\left. {1\over 4\kappa} \left(1-{k_{12}^2\over 4}\right) (f_u h_1 -g_v h_2 )\right\vert_{z_H(\tau_p)}^{z_H(\tau_f)},
\eeqa
both of which vanish for $h_1(v) =g_v(v)$ and $h_2(u) = f_u(u)$, so that the generalized Komar integral is conserved trivially and becomes zero in this case.

We thus consider a more non-trivial  case that $h_1(v)=c_1$ and $h_2(u)=c_2$ with constants $c_1,c_2$, and take $\tau_p < \tau_0 = {1\over \sqrt{2}}\min(1/a,1/b)$,
at which we have
\beqa
E_K(\tau_p) &=& -{1\over 4\kappa}  \left(1-{k_{12}^2\over 4}\right) \left[c_1 n_1a + c_2 n_2 b\right].
\eeqa
Note that $E_K(\tau_p)$ agrees with the energy of two plane gravitational waves before collision, eq.~\eqref{eq:E0}, if we take
$c_1=c_2=\displaystyle -{\sqrt{2}\over 1-k_{12}^2/4 }$.

Let us consider the following 3 cases for $\tau_f$ separately.
\begin{enumerate}
\item $\tau_f \le \tau_0$: In this case, the energy is conserved as $E_K(\tau_f) = E_K(\tau_p)$.
Indeed we also confirm $S_R(\tau_f,\tau_p)=S_L(\tau_f,\tau_p) =0$.
\item $\tau_0 < \tau_f \le \tau_1= {1\over \sqrt{2}}\max(1/a,1/b)$: In this case
\beqa
E(\tau_f) &=& {1\over 4\kappa}  \left(1-{k_{12}^2\over 4}\right)\times
\left\{
\begin{array}{ll}
c_1 f_u(u_R) -c_2\{ g_v(v_R)) + n_2 b\},  &   a\ge b ,  \\
\\
c_2 g_v(v_L) -c_1 \{ f_u(u_L)) + n_1 a\}, &     a< b, \\
\end{array}
\right. ,
\eeqa
where $u_H ={1\over\sqrt{2}}( \tau_f + z_H(\tau_f)$ and  $v_H ={1\over\sqrt{2}}( \tau_f - z_H(\tau_f)$ for $H=R,L$.
Since $E_K(\tau_f)\not=E_K(\tau_p)$, the generalized Komar integral is not conserved.
Indeed $E_K(\tau_f)- E_K(\tau_p)= S_R(\tau_f,\tau_p)-  S_L(\tau_f,\tau_p)\not=0$ is satisfied, where
\beqa
S_R(\tau_f,\tau_p) &=&\theta(a-b)  {1\over 4\kappa}  \left(1-{k_{12}^2\over 4}\right)\left[c_1\{ f(u_R) + n_1 a\} - c_2 g_v(v_R) \right], \\
S_L(\tau_f,\tau_p) &=&\theta(b-a)  {1\over 4\kappa}  \left(1-{k_{12}^2\over 4}\right)\left[-c_2\{ g_v(v_L) + n_2 b\} + c_1 f_u(u_L) \right].
\eeqa
\item $\tau_1< \tau_f\le \tau_{\rm end}$: In this case
\beqa
E(\tau_f) &=& {1\over 4\kappa}  \left(1-{k_{12}^2\over 4}\right)\left[ c_1 \{ f_u(u_R) - f_u(u_L)\} -c_2\{g_v(v_R) - g_v(v_L)\}\right].
\eeqa
The generalized Komar integral is not conserved as $E(\tau_f) - E(\tau_p) = S_R(\tau_f,\tau_p) - S_L(\tau_f,\tau_p) \not=0$, where
\beqa
S_R(\tau_f,\tau_p) &=&  {1\over 4\kappa}  \left(1-{k_{12}^2\over 4}\right)\left[c_1\{ f(u_R) + n_1 a\} - c_2 g_v(v_R) \right], \\
S_L(\tau_f,\tau_p) &=& {1\over 4\kappa}  \left(1-{k_{12}^2\over 4}\right)\left[-c_2\{ g_v(v_L) + n_2 b\} + c_1 f_u(u_L) \right].
\eeqa
Note that $E(\tau_{\rm end}) =0$ since $u_R=u_L$ and $v_R=v_L$.
\end{enumerate}

Let us conclude that, while the Komar current $K^\mu[\xi]$ is always conserved locally, the generalized Komar integral $E_K(\tau)$ is not conserved in general,
due to non-zero contributions from integrals on other boundaries, $S_R(\tau_f,\tau_p) - S_L(\tau_f,\tau_p)$.

It is worth mentioning that, as discussed in the main text,
integrals of $T^\mu{}_\nu \zeta^\nu$ on boundaries at $t^2=0$
are always zero,
since $ [d^3x]_\mu   T^\mu{}_\nu =0$ on these boundaries.
This is not an accidental, as any matters never flow into regions outside the defined spacetime.  
If some matter exists, the corresponding spacetime must also exist according to Einstein equation. 
On the other hand, this argument can not be applied to the Komar current since it does not have such a physical meaning,
 so that a conservation of the generalized Komar integral is not automatically guaranteed in general, as seen in this appendix.


\end{document}